\documentclass[reprint,aps,prb,superscriptaddress]{revtex4-2}

\usepackage{amsmath, amssymb, times, mathrsfs, hyperref, array, bbm}
\usepackage{graphicx}
\usepackage[usenames,dvipsnames]{xcolor}
\usepackage{braket}
\usepackage{xcolor}
\usepackage{tcolorbox}
\usepackage{listings}
\hypersetup{colorlinks=false}
\usepackage{hyperref}

\bibpunct{[}{]}{,}{n}{}{}
\usepackage{ulem}
\usepackage{lipsum}
\usepackage{listings}

\usepackage{mdframed}
\usepackage{balance}

\begin{document}
\title{Comment on ``Interferometric single-shot parity measurement in InAs–Al hybrid devices",\\
Microsoft Quantum, Nature 638, 651–655 (2025)}

\author{Henry F. Legg}
\affiliation{SUPA, School of Physics and Astronomy, University of St Andrews, North Haugh, St Andrews, KY16 9SS, United Kingdom}
\affiliation{Department of Physics, University of Basel, Klingelbergstrasse 82, CH-4056 Basel, Switzerland}

\begin{abstract}
We consider the `parity readout' of a (topological) superconductor claimed in Nature~638,~651–655~(2025). A prerequisite for this claim is the existence of a superconducting gap in the nanowire device. However, to determine the presence of a gap, Nature~638,~651–655~(2025) relied on the so-called topological gap protocol (TGP). Here, we show that the TGP can report the regions where the `parity readout' occurred as either gapped or gapless, depending on data parameters such as magnetic field range and cutter pair (junction transparency). Compounding these issues are inaccuracies in the presented TGP outcomes, which limited investigation of reproducibility. Since these inconsistent outcomes demonstrate that the TGP is not a reliable diagnostic tool for the presence of a superconducting gap, we instead investigate the conductance data for the regions where `parity readout' occurred --- data that were not presented in Nature~638,~651–655~(2025), but are in the public data repository. These conductance data show that these regions are in fact highly disordered and there is no clear superconducting gap in the nanowire, i.e., the underlying conductance data show that these regions are indeed gapless. That these regions are gapless contradicts the claim that the reported measurements are of the parity of a superconducting nanowire, let alone the parity of a topological superconducting nanowire. Taken together, these issues mean that the core findings in Nature~638,~651–655~(2025) are not reliable and should be revisited.
\end{abstract}
\maketitle

\onecolumngrid
\vspace{-34pt}
\section*{Summary of issues in Nature 638, 651–655 (2025), Microsoft Quantum (Ref.~1)}
\vspace{-10pt}
\begin{enumerate}
\itemsep0em 
     \item \textbf{The TGP can report the region investigated for `parity readout' as both gapped and gapless.} In Ref.~\citenum{Aghaee2025} the so-called topological gap protocol (TGP) is used to tune up the devices into a `gapped' (topological) phase. However, it was recently noted~\cite{Legg2025} that the outcome of the TGP is sensitive to data parameters such as data ranges and resolutions. We demonstrate that the main region studied in Ref.~\citenum{Aghaee2025} can be declared either gapped or gapless by the TGP, {\it e.g.}, depending on magnetic field range or chosen cutter pair. This emphasises that the TGP is not a reliable diagnostic tool for a gap.
\item \textbf{TGP outcomes not correct, inhibiting exploration of reproducibility.} It was claimed by the authors of Ref.~\citenum{Aghaee2025} that: ``In each measurement, this was the only region passing the TGP within the explored gate voltage and magnetic field range"~\cite{Aghaee2025peer}. This claim is incorrect. Even for the TGP exactly as applied in Ref.~\citenum{Aghaee2025}, Device B passed in several regions. This issue is further compounded by asymmetric bias-voltage values in the underlying data sets that cause an antisymmetrisation that is not around zero-bias, resulting in further large changes in the reported TGP outcomes of Ref.~\citenum{Aghaee2025}. Most importantly, this means that the `parity readout', if reproducible, should have been possible in multiple regions that passed the TGP, but such reproduction was not performed in Ref.~\citenum{Aghaee2025}. As such, it remains unanswered why the presented regions in Ref.~\citenum{Aghaee2025} were chosen and whether the effect is reproducible elsewhere in phase space.
\item \textbf{Conductance data that was not presented shows high levels of disorder and no clear superconducting gap.} In Ref.~\citenum{Aghaee2025} only TGP phase `cartoons' are presented. The true conductance data reveal that the regions of phase space where `readout' measurements were carried out are highly disordered and do not present a well-defined superconducting gap. Furthermore, it was stated: ``We have not detected the formation of accidental quantum dots in the junction region adjacent to the wire, as such dots would typically induce particle-hole symmetry breaking in the dc transport data, which is absent in our observations.'' However, this is not correct and the conductance data reveal considerable particle-hole symmetry breaking. These conductance data, not presented in Ref.~\citenum{Aghaee2025}, therefore strongly suggest that, even if reproducible, the effect reported in Ref.~\citenum{Aghaee2025} are due to  fine-tuned mesoscopic disorder physics, rather than a superconducting nanowire.
\end{enumerate}
\vspace{0pt}
\vspace{-5pt}
\twocolumngrid

\begin{figure*}[t]
  \includegraphics[width=1\textwidth]{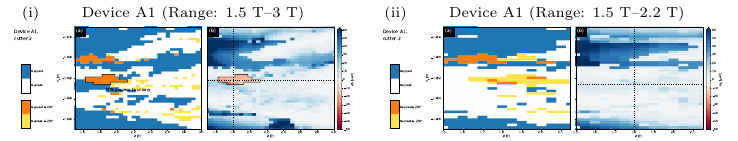}
	\caption{{\bf Readout region in Device A1 reported as both gapped and gapless depending on magnetic field range:} (i) The TGP phase cartoon shown in Ref.~\citenum{Aghaee2025} for Device A measurement 1, cutter 2. For this magnetic field range the device passes the TGP and is reported as gapped in the orange highlighted region. (ii) The outcome of the TGP for Device A measurement 1 for a reduced magnetic field range, the device now fails the TGP and the previous region that passed is seen as gapless (yellow, white).  This is despite the fact that this range still includes the region at $B=1.8$~T (vertical dashed lines) where `parity readout' occurs. The horizontal dashed line is the plunger cut shown in Fig.~\ref{A1_con}, but it should be noted that the exact corresponding plunger voltage range used for this `readout' is unclear due to "hysteresis" and "cross capacitance"~\cite{Aghaee2025peer}. {\sl Note: We have chosen not to change label sizes throughout this Comment in order to make minimal alterations to the TGP code.}}
\label{A_TGP_Outcome}
\vspace{-5pt}
\end{figure*}
\begin{figure*}[t]
  \includegraphics[width=1\textwidth]{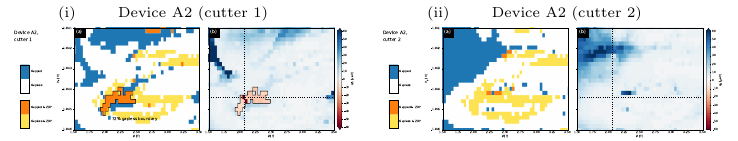}
	\caption{{\bf Readout region in Device A2 reported as both gapped and gapless depending on cutter pair:} (i) The TGP phase cartoon shown in Ref.~\citenum{Aghaee2025} for Device A measurement 2, cutter 1. (ii) The outcome of the TGP for Device A measurement 2 , cutter 2. The TGP reports the same orange highlighted region in (i) as gapless (yellow) for the cutter in (ii). This is despite the fact that both phase cartoons still include the region at $B=2.06$~T where `parity readout’ occurs (vertical dashed line). The horizontal dashed lines are the plunger cuts shown in Fig.~\ref{A2_con}.}
\label{A2_TGP_Outcome}
\vspace{-5pt}
\end{figure*}

{\bf Overview.} In Nature 638, 651–655 (2025)~\cite{Aghaee2025} Microsoft Quantum purport to measure the parity --- whether total number of fermions is even or odd --- of a proximitised superconducting nanowire. Whilst the authors imply throughout Ref.~\citenum{Aghaee2025} that the measurement is of a topological superconductor, they acknowledge the reported effect could be due to a trivial superconductor. However, for the parity of the proximitised nanowire to be protected from low-energy quasiparticles, the presence of a superconducting (SC) gap is required~\cite{Rainis2012,Albrecht2017,Karzig2017}.

In this comment, using the public data~\cite{Github} and peer review file~\cite{Aghaee2025peer} for Ref.~\citenum{Aghaee2025}, we show that the tune-up method used in Ref.~\citenum{Aghaee2025} to detect a gap --- the topological gap protocol (TGP) --- declares the regions of phase space where `parity readout' occurred both gapped and gapless, depending on data parameters. These issues with the TGP are compounded by incorrect outcomes presented in Refs.~\citenum{Aghaee2025} and coding issues, which limited investigations of the reproducibility of the effect. 

Given these inconsistent TGP outcomes, we investigate the underlying conductance data, not presented in Ref.~\citenum{Aghaee2025}. These data reveal that there is no clear signature of an SC gap and that the nanowire is highly disordered with many low-energy states. This strongly suggests that the purported `readout' of the nanowire, if reproducible, is a fine-tuned mesoscopic disorder effect, rather than due to well-defined superconductivity in the nanowire. As such, the core claims of Ref.~\citenum{Aghaee2025} are not reliable and should be revisited.

\section{Inconsistent TGP gap outcomes\\ for regions of `parity readout'}
\vspace{-10pt}
The TGP \cite{Agahee2023} is a purportedly ``stringent'' test whether a SC gap occurs in a nanowire device and, if so, whether the region where that gap occurs is topological. In Ref.~\citenum{Aghaee2025} the TGP is used to tune up the studied nanowire devices into a supposedly gapped (topological) superconducting phase. In particular, it is used to select the gate range and magnetic field setting used for the main effect reported in Ref.~\citenum{Aghaee2025}.

However, it was recently noted in Ref.~\citenum{Legg2025} that the TGP outcome is sensitive to the ranges and resolutions of the underlying datasets (e.g., magnetic field and bias voltage range), as well as to the choice of cutter pair (junction transparency). Consequently, changes in these parameters can alter whether a region of parameter space is seen as gapped or gapless by the TGP and ultimately whether it passes or fails the TGP. Since Ref.~\citenum{Aghaee2025} does not conclusively claim to have measured a topological superconductor, in this comment we will not focus on whether the region is identified as topological by the TGP, but rather simply whether the region is reported as gapped or gapless. Ultimately this is the key question for claim of Ref.~\citenum{Aghaee2025}, because if the system does not have a well-defined SC gap then low-energy quasiparticles can easily switch parity~\cite{Rainis2012,Albrecht2017,Karzig2017}.

First, in Fig.~\ref{A_TGP_Outcome} we show that the main region studied in Device A measurement 1 fails the TGP and is reported as gapless for a different choice of magnetic field range (1.5 T -- 2.2 T) than in Ref.~\citenum{Aghaee2025} that still includes the region at $B=1.8$~T where the purported parity readout occurs. This is the same TGP inconsistency as was reported in Ref.~\citenum{Legg2025}, but now also shown for the system studied in Ref.~\citenum{Aghaee2025}.

As was also reported in Ref.~\citenum{Legg2025}, the chosen cutter pair (junction transparency) can also change whether the TGP sees the identified region of phase space as gapped or gapless. In Fig.~\ref{A2_TGP_Outcome} we show that the main region studied in Device A measurement 2 ($B=2.06$~T) is reported as gapless for cutter number 2. The reason why the TGP reports these regions as either gapped or gapless is because there is still sufficient nonlocal conductance at low-bias that changing the magnetic field range and/or cutter can alter whether the nonlocal conductance surpasses the threshold of 5\% of the maximum that is required for the TGP to see the region as gapped~\cite{Agahee2023,Legg2025}. 

Altogether, these inconsistent outcomes for whether the TGP reports the same region as gapped or gapless reinforces that the TGP does not provide a reliable means for detecting a (topological) superconducting gap and so the tune-up procedure in Ref.~\cite{Aghaee2025} is not a reliable method to identified a gapped region of phase where parity should be protected by a superconducting gap. Finally, it should also be noted, as was observed for Ref.~\citenum{Agahee2023}, there are variations in the data parameters of the underlying datasets used for Ref.~\citenum{Aghaee2025}, even for the same device. The bias range is also small, for example, in Device~A1 it runs from \mbox{-66.18\,$\mu$V} to \mbox{+63.64\,$\mu$V}. Note also that this range is not symmetric around zero-bias.
\vspace{0pt}

\begin{figure}[t]
  \includegraphics[width=1\columnwidth]{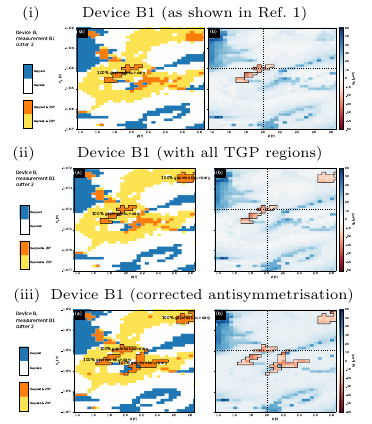}
	\caption{{\bf TGP for Device B:} {\bf (i)} The TGP phase cartoon shown in Ref.~\citenum{Aghaee2025}. Only the largest region passing the TGP is shown due to the use of {\sl zbp\_cluster\_numbers=[1]} in the plotting function of the TGP outcome. The vertical dashed line shows the region at $B=2.0$~T where `readout' ocucrred. {\bf (ii)} The code {\sl zbp\_cluster\_numbers=[1,2]} reveals that a second region [orange highlighted, red in (b)] around $B\approx2.8$~T actually also passed the TGP, but was not shown or investigated in Ref.~\citenum{Aghaee2025}. This second region passing the TGP contrasts with what was claimed during the review process~\cite{Aghaee2025peer}. {\bf (iii)} Due to the asymmetric bias voltages of supplied data sets the TGP does not correctly antisymmetrise around zero-bias. If instead this single extra bias value is remove such that the bias range is symmetric, we see that, in comparison to Fig.~\ref{B_TGP_Outcome} the TGP is passed in even more regions of phase space and the explored region for `parity readout' around $B=2.0$~T is actually secondary to larger regions of phase space that are `gapped' and `topological' at high magnetic field but were not investigated. This shows that the large impact even a single bias voltage point can have on the TGP outcome.}
\label{B_TGP_Outcome}
\end{figure}

\vspace{-10pt}
\section{Reported TGP outcomes not correct}
Not only does the TGP give inconsistent results, but the TGP outcomes reported in Ref.~\citenum{Aghaee2025} were also not accurate. In the open peer review file (Ref.~\citenum{Aghaee2025peer}) several reviewers queried whether further regions pass the TGP --- where one would expect the reported `parity readout' to be reproducible. For instance, it was asked~\citenum{Aghaee2025peer}: 
\begin{quote}
``Are A1/A2 the only regions that passed TGP? Are there other regions where things didn’t work out
but, based on the TGP, should have? Similar questions for Device B. The authors need to give us some context to make the control experiments meaningful.''
\end{quote}
In response to these queries the authors of Ref.~\citenum{Aghaee2025} claimed~\cite{Aghaee2025peer}:  
\begin{quote}
``In each measurement, this was the {\bf only} region passing the TGP within the explored gate voltage and magnetic field range.''
\end{quote}
This statement is incorrect. As shown in Fig.~\ref{B_TGP_Outcome}(ii), applying the TGP as implemented in Ref.~\citenum{Aghaee2025} but now with all regions identified by the TGP shown, Device~B exhibits multiple regions that are identified by the TGP. In particular, in Fig.~\ref{B_TGP_Outcome} there is a region ($V_{\rm p}\approx-1.6725$~V, $B\approx 2.8$~T) that is identified by the TGP and where no `parity readout' data was presented (a further third region in this device is present for other cutters and also passes the TGP). The only reason this region was not identified in the TGP phase cartoon shown in Ref.~\citenum{Aghaee2025} was because the input to the TGP plotting code caused  only the largest identified region to be highlighted. 

Since this region of phase space was identified by the TGP, the purported parity readout should have been reproducible there, but this is not investigated in Ref.~\citenum{Aghaee2025}. As such, this incorrect reporting of the TGP outcomes strongly limited exploration of the expected reproducibility of the main effect reported in Ref.~\citenum{Aghaee2025}, despite reproducibility in other regions being a central question asked by several referees.

This error in reporting the TGP outcome for Device B is further compounded by a coding issue in the antisymmetrization of the nonlocal conductance data by the TGP. Namely, the code antisymmetrises based on Python array index rather than bias voltage (see code extract). As noted above, the bias-voltages for the provided data in Ref.~\citenum{Aghaee2025} are not symmetric around zero leading to antisymmetrisation that is not performed around zero-bias. As shown in Fig.~\ref{B_TGP_Outcome}(iii), the TGP outcome with a symmetric bias range (dropping the first value in the Python array) reveals that several larger regions pass the TGP in Device B. The fact that the focus in Ref.~\citenum{Aghaee2025} is actually on a small, secondary magnetic field value [vertical dashed line in Fig.~\ref{B_TGP_Outcome}(iii)], further underscores the disconnect between the regions where purported parity readout should have been possible and what was presented in Ref.~\citenum{Aghaee2025}. This also emphasises the large impact a single data entry in the bias-voltage can have on whether a region is identified by the TGP, further demonstrating that the TGP is not a reliable measure of an SC gapped (topological) superconducting regions of phase space.

\vspace{10pt}
{\footnotesize \begin{tcolorbox}[colback=gray!20, colframe=black, sharp corners, boxrule=0.1pt, width=\columnwidth]
\centering
\begin{verbatim}
999 def antisymmetric_conductance_part(
...
1007      lambda x: (x - x[::-1]) / 2,
\end{verbatim}
\end{tcolorbox}}
\vspace{-5pt}
\noindent {\bf Code extract showing antisymmetrisation is by array index not bias-voltage:} Code extract from \href{https://github.com/microsoft/azure-quantum-tgp/blob/main/tgp/two.py}{Two.py}. The way this antisymmetrisation is defined results in the TGP incorrectly antisymmetrising asymmetric data, as is the case for the released datasets of Ref.~\citenum{Aghaee2025}. As a consequence the TGP phase diagrams reported in Ref.~\citenum{Aghaee2025} are not correctly antisymmetrised. 

\begin{figure*}[t]
  \includegraphics[width=0.8\textwidth]{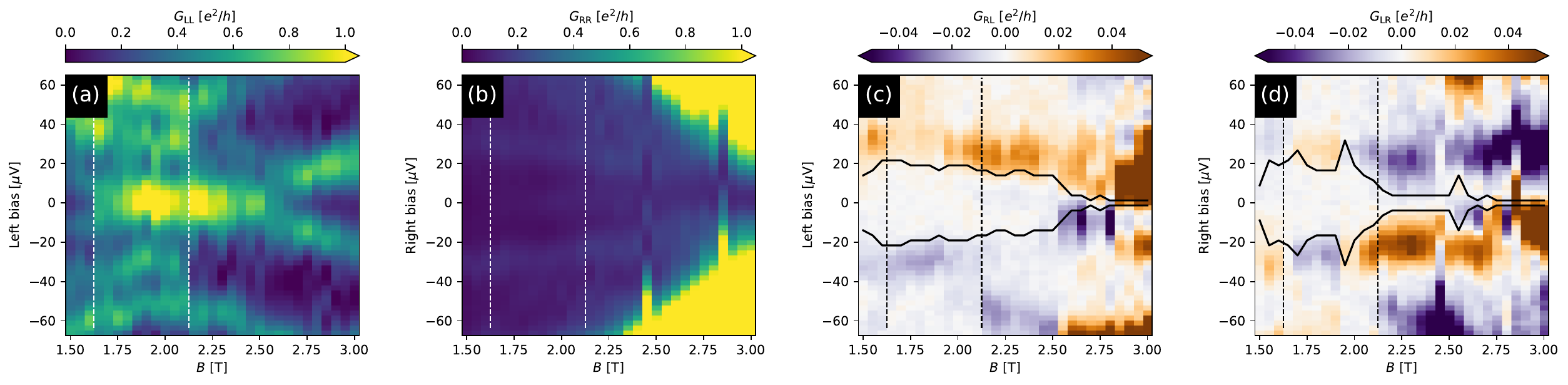}
	\caption{{\bf Conductance plots for Device A1.} Cuts along the horizontal dashed line through the identified region in Fig.~\ref{A_TGP_Outcome}. Note: No conductance plots are shown in Ref.~\citenum{Aghaee2025}. The local conductance (a-b) reveals a substantial number of low energy states on the left junction, with a much weaker signal on the right, but with several low-energy states visible. There is also large enough nonlocal conductance (c-d) at low-bias in the identified region for that region to reported as gapless when the magnetic field range is adjusted.}
\vspace{-5pt}
\label{A1_con}
\end{figure*}
\begin{figure*}[t]
  \includegraphics[width=0.84\textwidth]{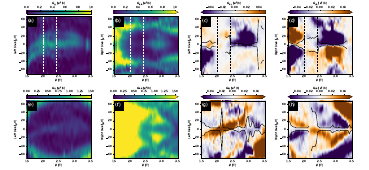}
	\caption{{\bf Conductance plots for Device A2.} Cuts along the horizontal dashed lines through the identified region in Fig.~\ref{A2_TGP_Outcome} for cutter 1 (a-d) and cutter 2 (e-h). Note: No conductance plots are shown in Ref.~\citenum{Aghaee2025}. The local conductance reveals a substantial number of low-energy states, with finite $G_{\rm RR}$ in (f) for almost all bias. There is a considerable particle-hole symmetry breaking, in contrast to the claim in Ref.~\citenum{Aghaee2025}, and a very substantial nonlocal conductance at low-bias . The solid black lines show the gap reported by the TGP. It should be noted that the antisymmetrisation in bias voltage of the TGP leads to this reported gap being insensitive to symmetric components of the conductance which is why a `gap' can be reported in the presence of large low-bias conductance. As above, this device is seen as gapped for cutter 1, but gapless for cutter 2 for the region where `parity readout' occurs (dashed black vertical lines indicate the region identified by the TGP).}
\label{A2_con}
\vspace{-12pt}
\end{figure*}

\begin{figure*}[t]
  \includegraphics[width=0.84\textwidth]{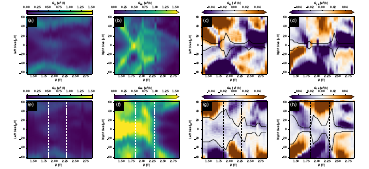}
	\caption{{\bf Conductance plots for Device B1.} Cuts along the horizontal dashed lines through the identified region in Fig.~\ref{B_TGP_Outcome}(iii) for cutter 1 (a-d) and cutter 2 (e-h) of Device B1. Note: No conductance plots are shown in Ref.~\citenum{Aghaee2025}. As in the other devices, this data reveals a substantial number of low-energy states, particle-hole symmetry breaking, and large nonlocal conductance at low-bias. It should be noted for this cut the system is seen as gapless for cutter 1, but gapped for cutter 2 (dashed vertical lines indicate region identified by the TGP).}
	\vspace{-13pt}
\label{B1_con}
\end{figure*}

\section{Conductance data not presented shows high levels of disorder and no SC gap}
The inconsistent and incorrect TGP outcomes for the data of Ref.~\citenum{Aghaee2025} further evinces that the TGP is not a reliable diagnostic tool for the existence of an SC gap in the nanowire. As such, we instead investigate the actual conductance data for each device. It should be emphasised that these data were not presented in Ref.~\citenum{Aghaee2025}, even for the specific parameter regions that passed the TGP. Only TGP phase `cartoons' are presented in the supplemental material of Ref.~\citenum{Aghaee2025}. These conductance data reveal that the regions where `parity readout' occur are actually highly disordered and do not present features consistent with a well-defined superconducting gap.

Starting with the local conductance in Figs.~\ref{A1_con}, \ref{A2_con}, \ref{B1_con}, it can be seen that there are a very large number of low energy states. Although the case for all devices, this is particularly striking for Device A2, cutter 2 [Fig.~\ref{A2_TGP_Outcome}(f)], where there is strong finite local conductance, $G_{\rm RR}$, for almost the full range of bias voltages, indicating an abundance of low-energy states and there is also no clear SC gap features. It is also interesting to note that in all cases there is a sizable difference in magnitude of the local signal on each side of the nanowire.

Similarly, the nonlocal conductance data in Figs.~\ref{A1_con}, \ref{A2_con}, \ref{B1_con} reveal a large level of disorder and significant nonlocal conductance at low-bias consistent with a gapless system, as indicated by some of the TGP outcomes. For instance, in Device A2 both cutters shown in Fig.~\ref{A2_TGP_Outcome} have very large amounts of nonlocal conductance close to zero-bias. 

Furthermore, it was claimed~\cite{Aghaee2025peer} by the authors of Ref.~\cite{Aghaee2025}:
\begin{quote}
``We have not detected the formation of accidental quantum dots in the junction region adjacent to the wire, as such dots would typically induce particle-hole symmetry breaking in the dc transport data, which is absent in our observations.'' 
\end{quote}
However, as can be seen, the full conductance data contradict this statement with both the local conductance and nonlocal conductance show broken particle-hole symmetry. In particular, considerable antisymmetric components of local conductance and symmetric components of nonlocal conductance are extremely clear, for instance, in Device A2 Fig.~\ref{A2_con}. Crucially, while the full nonlocal conductance can be substantial at low-bias and highly symmetric around zero-bias, the TGP is not sensitive to this due to the antisymmetrising of the conductance data, effectively cancelling out the symmetric component. Note that throughout we plot the full nonlocal conductance (not antisymmetrised) and, as a result, the solid black “gap” lines in the TGP plots (Figs. \ref{A1_con}, \ref{A2_con}, \ref{B1_con}) do not reflect what the TGP inputs rather than the full nonlocal conductance.

This investigation of the local and nonlocal conductance data therefore demonstrates that the TGP outcomes reporting the devices as gapless were (more) accurate and there is indeed no clear superconducting gap in the regions where `parity readout' occured. This is important because a system without a well-defined SC gap is not protected from low-energy quasiparticles altering the parity and this is inconsistent with the claims of Ref~\citenum{Aghaee2025}. It also reveals that these systems are highly disordered in the region where `parity readout' occurs, with a sizable particle-hole symmetry breaking component of the conductance data. 

Taken together, these conductance data reveal that the devices studied in Ref.~\citenum{Aghaee2025} are highly disordered and have no clear superconducting gap. This strongly suggests that the `parity readout' observed in Ref.~\citenum{Aghaee2025} actually arises from a fine-tuned mesoscopic disorder effect rather than a (topological) superconducting phase in the nanowire. That such a large level of disorder existed in these devices and that there was no clear SC gap was obscured by the fact only TGP phase cartoons are shown in Ref.~\citenum{Aghaee2025}, rather than conductance data, and further obscured due to an inaccurate descriptions stating that particle-hole breaking is ``absent in our observations''.

\section*{Conclusions}
\vspace{-5pt}
As observed in Ref.~\citenum{Legg2025} for the first devices `passing' the TGP [PRB 107, 245423 (2023)], our analysis of Nature 638, 651–655 (2025)~\cite{Aghaee2025} indicates that the TGP is also not a reliable diagnostic tool for the devices and effect studied in Ref.~\citenum{Aghaee2025}. It should be emphasised that this unreliability is not only for the detection of topology, {\it i.e.} Majorana bound states, but also for the detection of a gap in these devices. In particular, in Ref.~\citenum{Aghaee2025}, the TGP is used to identify purportedly gapped (topological) superconducting phases in Ref.~\citenum{Aghaee2025}, but its outcome is dependent on chosen data parameters (ranges, resolutions, and cutter settings). This is important because the TGP does not provide a reliable diagnostic for whether a region is gapped or gapless and this is a prerequisite for the parity of the superconducting nanowire to be protected. Furthermore, there were additional regions satisfying the TGP that were not investigated in Ref.~\citenum{Aghaee2025} due to incorrect TGP outcomes in Ref.~\citenum{Aghaee2025}. This means that reproduction of the reported effect could not be properly established in other regions of phase space where it would be expected to occur.

Most fundamentally, since the TGP is not a reliable diagnostic tool for an SC gap, the full conductance data in the regions where `parity readout' measurements occurred are the only way to establish if a gap is present. These data clearly show an abundance of low-energy states, strong particle-hole symmetry breaking, and no clear superconducting gap. As such, these data strongly suggest that the effect reported in Ref.~\citenum{Aghaee2025} is a fine-tuned mesoscopic disorder effect and show that it cannot be related to a (topological) superconducting nanowire due to the lack of a well-defined gap in the nanowire. However, by only presenting TGP phase cartoons the true nature of the conductance data in Ref.~\citenum{Aghaee2025} was obscured. Taken together, in particular since the actual conductance data are not shown in Ref.~\citenum{Aghaee2025}, these issues demonstrate that the findings of Ref.~\citenum{Aghaee2025} are not reliable and should be revisited.
\newpage
\onecolumngrid
The code to reproduce the figures in this comment are available under in the Zenodo repository: \href{https://doi.org/10.5281/zenodo.15008728}{10.5281/zenodo.15008728}
\twocolumngrid
\bibliography{refs}
\end{document}